\documentclass[10pt]{article}

\usepackage{url} 
\usepackage{float}
\usepackage{graphicx}
\usepackage{amsfonts, amssymb, amscd}
\usepackage{amsmath, delarray, theorem}
\usepackage{natbib}
\bibpunct{(}{)}{;}{a}{}{,}
\renewcommand{\cite}{\citep}

\newcommand{\bitem}{\begin{itemize}}
\newcommand{\eitem}{\end{itemize}}
\newcommand{\benum}{\begin{enumerate}}
\newcommand{\eenum}{\end{enumerate}}
\newcommand{\beqnn}{\begin{eqnarray*}}
\newcommand{\eeqnn}{\end{eqnarray*}}
\newcommand{\beqn}{\begin{eqnarray}}
\newcommand{\eeqn}{\end{eqnarray}}
\newcommand{\normal}[2]{\mbox{N}(#1,#2)}

\newcommand{\var}{\mbox{Var}}

\newcommand{\corr}{\mbox{Corr}}

\newcommand{\myvec}[1]{\mathbf{#1}}

\newcommand{\bfourmatrix}{\begin{array}({cccc})}
\newcommand{\efourmatrix}{\end{array}}
\newcommand{\bvector}{\begin{array}({c})}
\newcommand{\evector}{\end{array}}

\newcommand{\btab}{\renewcommand{\baselinestretch}{1}\begin{table}[hptb]}
\newcommand{\etab}{\end{table}\renewcommand{\baselinestretch}{1}}
\newcommand{\bfig}{\renewcommand{\baselinestretch}{1}\begin{figure}[hptb]}
\newcommand{\efig}{\end{figure}\renewcommand{\baselinestretch}{1}}

\newcommand{\gvec}[1]{\mbox{\boldmath ${#1}$}}

{\theoremstyle{plain}
  }
{\theoremstyle{plain}
  }
{\theoremstyle{plain}
  }
{\theoremstyle{plain}
  }
{\theorembodyfont{\rmfamily}
  }
{\theorembodyfont{\rmfamily}
  }

\renewcommand{\baselinestretch}{1.25}

\begin{document}
\bibliographystyle{natbib}

\title{Stochastic Stepwise Ensembles for Variable Selection}
\author{Lu Xin and Mu Zhu \\
University of Waterloo\\ 
Waterloo, ON, Canada N2L 3G1}
\date{\today}
\maketitle

\begin{abstract}
In this article, we advocate the ensemble approach for variable 
selection. We point out that the stochastic mechanism used to generate the 
variable-selection ensemble (VSE) must be picked with care. We construct a 
VSE using a stochastic stepwise algorithm, and compare its performance 
with numerous state-of-the-art algorithms.
\end{abstract}

\newpage

\section{Introduction}

The ensemble approach for statistical modelling was first made popular 
by such algorithms as boosting \citep{boosting-orig, boosting}, bagging 
\citep{bagging}, random forest \citep{randomForest}, and the gradient 
boosting machine \citep{gradboost}. They are powerful algorithms for 
solving prediction problems. This article is concerned with using the 
ensemble approach for a different problem, variable selection. We shall 
use the terms ``prediction ensemble'' and ``variable-selection ensemble'' 
to differentiate ensembles used for these different purposes.

\subsection{Variable-selection ensembles (VSEs)}
\label{sec:VSE}

First, we give a general description of variable-selection ensembles 
(VSEs).
Suppose there are $p$ candidate variables. A VSE (of size $B$) can be 
represented by a $B \times p$ matrix, say 
$\myvec{E}$, whose $j$-th column contains $B$ independent measures of how 
important variable $j$ is. Let $\myvec{E}(b,j)$ denote the $(b,j)$-th 
entry of $\myvec{E}$. Using the ensemble $\myvec{E}$ as a whole, one 
typically ranks the importance of variable $j$ using a majority-vote type 
of summary, such as 
\beqn
\label{eq:rank}
 R(j) = \frac{1}{B} \sum_{b=1}^B \myvec{E}(b,j),
\eeqn 
and the variables that are ranked ``considerably higher'' than the rest 
are then selected. 

The key for generating a VSE lies in producing multiple measures of 
importance for each candidate variable. By contrast, traditional variable 
selection procedures, including stepwise selection and Lasso, typically 
produce just one such measure, that is, $B=1$. It shouldn't be hard for 
any statistician to appreciate that averaging over a number of independent 
measures is often beneficial. This is the main reason why VSEs are 
attractive and more powerful than many traditional approaches.

To make selection decisions, one must be more precise about what it means 
to say that some variables are ranked ``considerably higher'' than the 
rest. 
One option is to select variable 
$j$ if it is ranked ``above average,'' i.e., if
\beqn
\label{eq:meanthresh}
R(j) > \frac{1}{p} \sum_{k=1}^p R(k).
\eeqn
This is what we use
in all the experiments reported below, but we 
emphasize that other thresholding rules can be used as well. 
For example, one can make 
a so-called ``scree plot'' of $R(1), R(2), ..., R(p)$, and look for an 
``elbow'' --- a very common practice in principal component analysis 
\citep[e.g.][]{jolliffe}, but the precise location of the 
``elbow'' is highly subjective, which is why we choose not to use this 
strategy here.

The distinction between ranking and thresholding is particularly important 
for VSEs; we will say more about this in Section~\ref{sec:rank-select}.

\subsection{PGA}
\label{sec:pga}

\citet{pga} constructed a VSE using a so-called parallel genetic algorithm 
(PGA). To produce multiple measures of variable importance, PGA repeatedly 
performs stochastic rather than deterministic optimization of the Akaike 
Information Criterion \citep[AIC;][]{aic}, while deliberately stopping 
each optimization path prematurely. In practice, one must be more exact 
about what ``premature'' means, but we won't go into the specifics here 
and refer the readers to \citet[][Section 3.2]{pga} for details.

The main idea is as follows. Early termination forces each optimization 
path to produce sub-optimal rather than optimal solutions, while the use 
of stochastic rather than deterministic optimization allows each of these 
sub-optimal solutions to be different from each other \citep{mz-tas2008}. 
For example, 
suppose we have five candidate variables, $x_1, x_2, ..., x_5$. The first 
time we stochastically optimize the AIC for just a few steps, we may 
arrive at the solution $\{x_1, x_2, x_3\}$; the second time, we may arrive 
at $\{x_1, x_2, x_4\}$; and the third time, perhaps $\{x_1, x_2, x_5\}$.
This produces the following ensemble:
\[
 \myvec{E} = \left[ 
 \begin{array}{ccccc}
 1 & 1 & 1 & 0 & 0 \\
 1 & 1 & 0 & 1 & 0 \\
 1 & 1 & 0 & 0 & 1 
 \end{array} \right].
\] 
Since
\[
 R(1) = R(2) = 1 > \frac{1}{3} = R(3) = R(4) = R(5),
\]
the ensemble selects $\{x_1, x_2\}$. 

\citet{pga} used the genetic algorithm \citep[GA;][]{gen-algo} as their 
stochastic optimizer in each path, but our general description of VSEs 
above (Section~\ref{sec:VSE}) makes it clear that any other stochastic 
optimizer can be used for PGA to work, despite the name ``PGA.'' This is 
a crucial point, and we will come back to it later 
(Sections~\ref{sec:ex0} and \ref{sec:bigpic}).

Though driven by the AIC, it has been observed that PGA has a much higher 
probability of selecting the correct subset of variables than optimizing 
the AIC by exhaustive search --- of course, such observations have only 
been made on mid-sized simulation problems where an exhaustive search is 
feasible and the correct subset is known. Nonetheless, they show quite 
conclusively that PGA is not merely a better search algorithm, because one 
cannot possibly perform a better search than an exhaustive one. 
Therefore, PGA can be seen as an effective AIC ``booster,'' and this is 
precisely why the ensemble approach to variable selection is 
valuable and powerful. 

\subsection{Motivating example: A weak signal} 
\label{sec:ex0}

One of the key objectives of this article is to present a better 
variable-selection ensemble than what PGA produces, one which we call the 
stochastic stepwise ensemble (ST2E). Like \citet{pga}, we also focus on 
multiple linear regression models, although VSEs are easily applicable to 
other statistical models such as logistic regression and Cox regression.

We first describe a simple experiment to motivate our work.
There are 20 potential 
predictors, $\myvec{x}_1, \myvec{x}_2, ..., \myvec{x}_{20}$, but only 
three of them are actually 
used in the true model to 
generate the response, $\myvec{y}$:
\beqn
\label{eq:ex0}
 \myvec{y} = \alpha \myvec{x}_{1} + 2\myvec{x}_{2} + 3\myvec{x}_{3} + 
 \sigma \gvec{\epsilon},
 \quad \myvec{x}_{1}, ..., \myvec{x}_{20}, \gvec{\epsilon}
 \sim
 \normal{\myvec{0}}{\myvec{I}}.
\eeqn
The sample size $n$ is taken to be $100$, and $\sigma=3$.
In addition, the three variables that generate $\myvec{y}$ are correlated,
with $\corr(\myvec{x}_i,\myvec{x}_j)=0.7$ for $i,j \in 
\{1,2,3\}$ and $i \neq j$. The $\myvec{x}$'s 
and $\gvec{\epsilon}$ are otherwise independent of 
each other. 

We shall consider 
$\alpha = 0.1, 0.2, ..., 0.9, 1.0$, $1.2$ and $1.5$. 
By construction, there are three types of variables: $\myvec{x}_1$ is a 
relatively weak variable --- it is part of the true model but its 
signal-to-noise ratio is relatively low; $\myvec{x}_2$ and 
$\myvec{x}_3$ are relatively strong variables; and $\myvec{x}_4, ..., 
\myvec{x}_{20}$ are noise variables --- they are not part of the true 
model.

\begin{figure}[hptb]
\centering
\includegraphics[width=0.325\textwidth, angle=270]{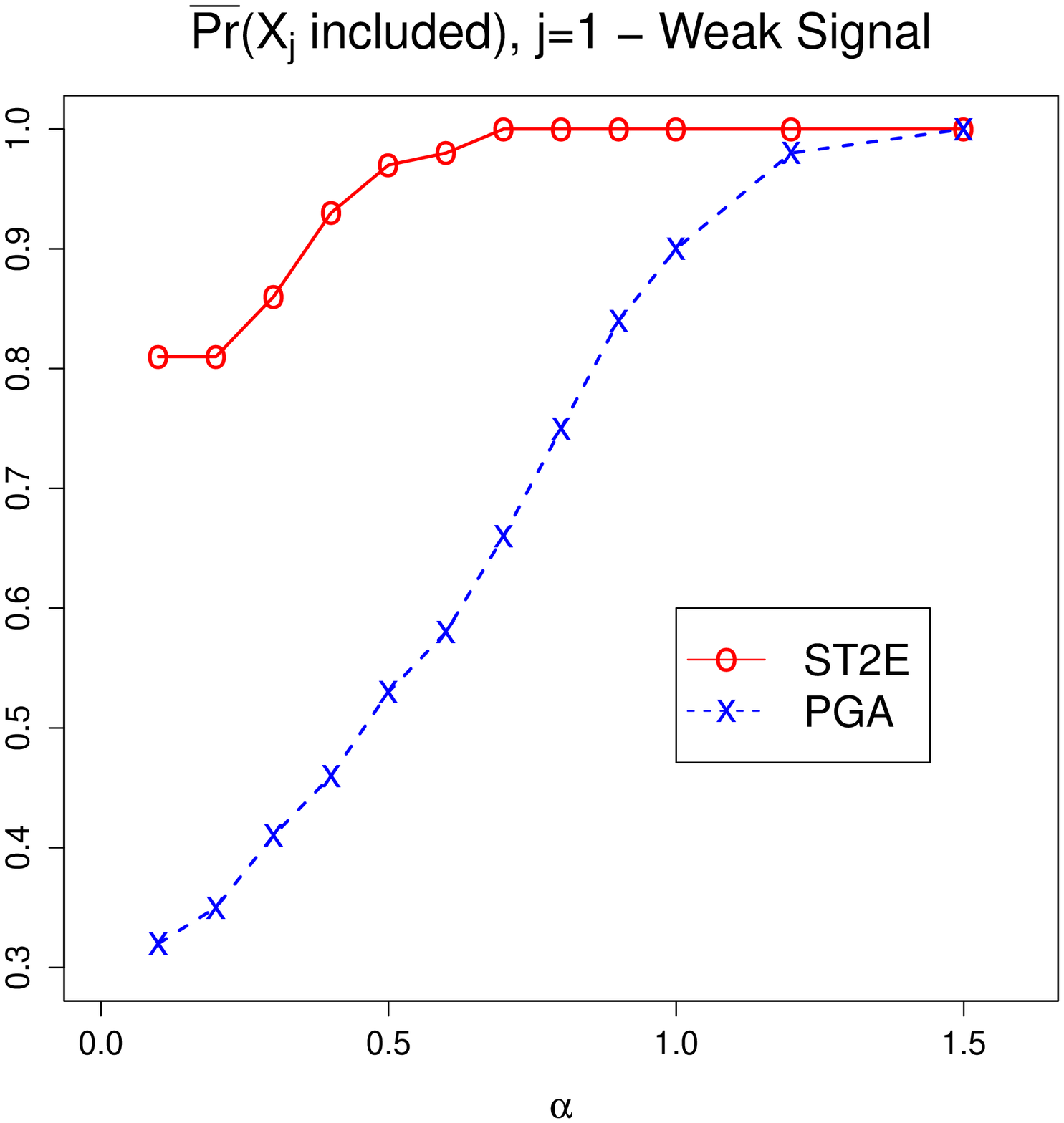}
\includegraphics[width=0.325\textwidth, angle=270]{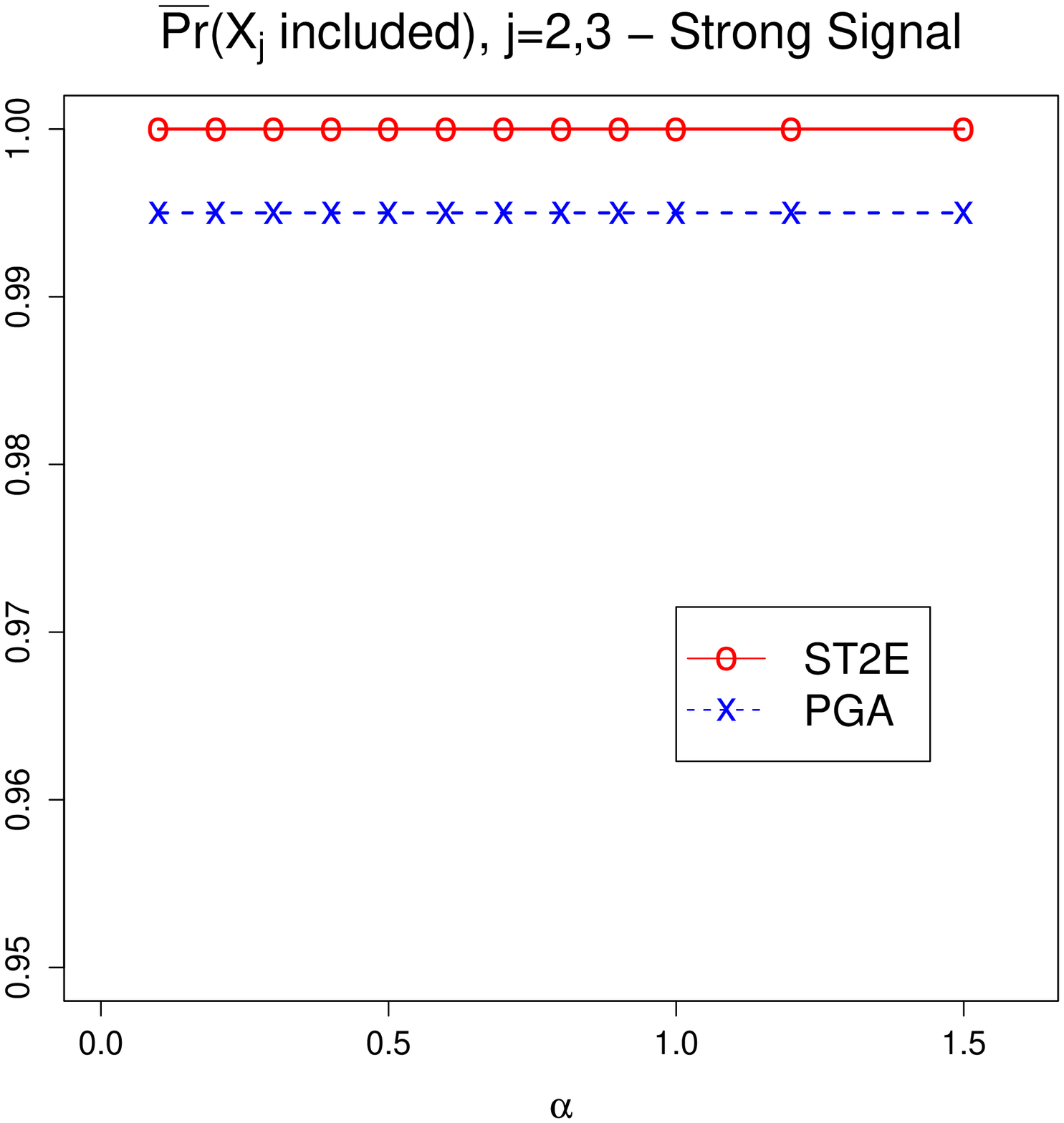}
\includegraphics[width=0.325\textwidth, angle=270]{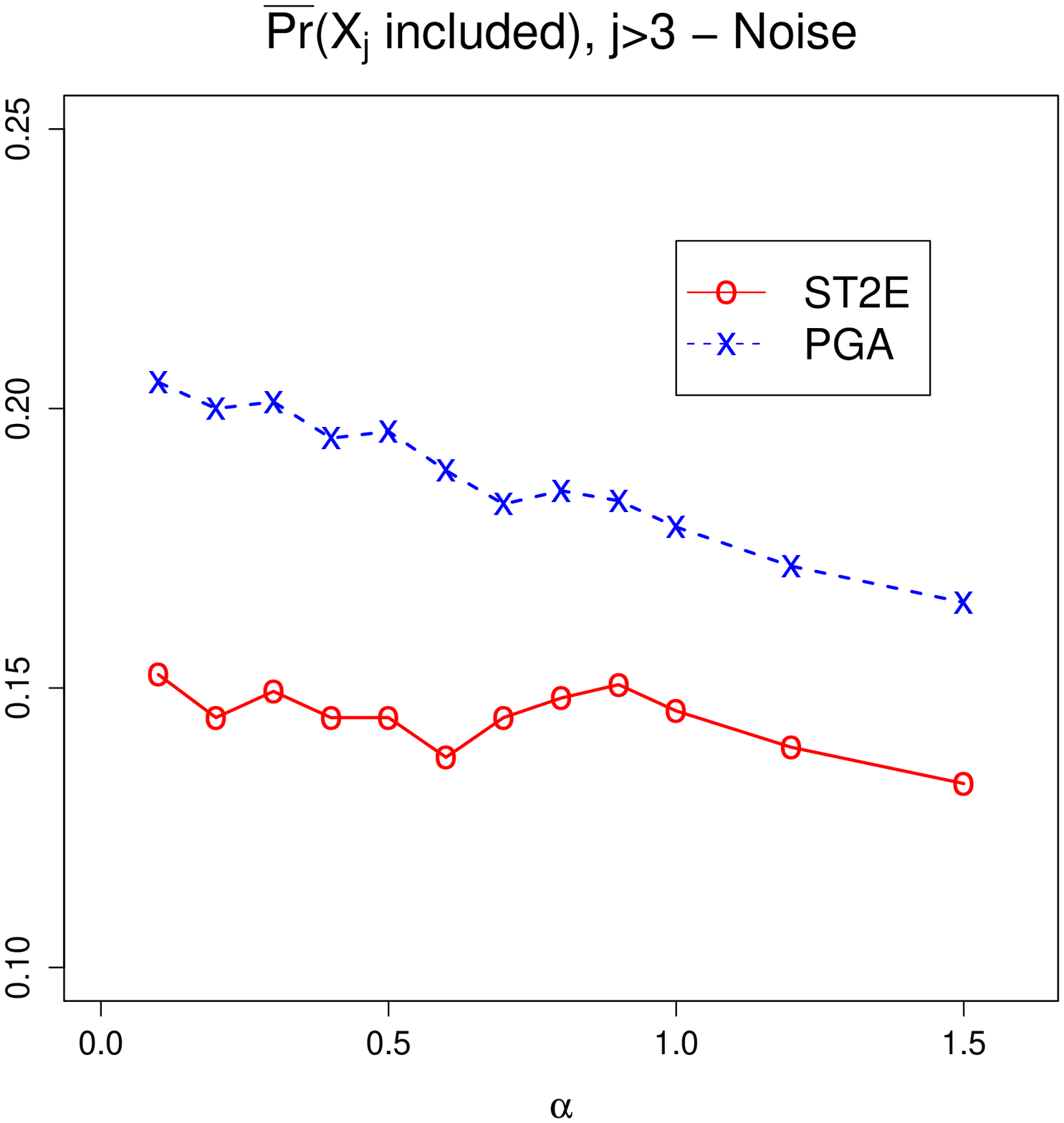}
\caption{\label{fig:pga-weakness}%
Motivating example (Section~\ref{sec:ex0}).
Average frequency the three different types 
of variables are selected by ST2E and PGA. Notice 
that the vertical scales are not identical.}
\end{figure}

For each $\alpha$, the experiment is repeated 100 times. 
Figure~\ref{fig:pga-weakness} shows the average 
frequency 
the three different types of variables are selected by two different 
VSEs, PGA and ST2E, as our experimental parameter $\alpha$ varies. 
The two VSEs are both of size $B=300$.

The messages from this experiment are as follows. In terms of catching the 
strong signals ($\myvec{x}_2$ and $\myvec{x}_3$), ST2E and PGA are about 
the same. In terms of guarding against the noise variables ($\myvec{x}_j$ 
for $j>3$), ST2E is slightly better than PGA. But, most importantly, we 
see that ST2E is significantly better than PGA at catching the weak 
signal, $\myvec{x}_1$. It is in this sense that ST2E is a better VSE than 
PGA. 

The improved performance of ST2E is due to the use of 
a more structured stochastic optimizer in each path. In particular, a 
so-called stochastic stepwise (STST or simply ST2) algorithm is used, 
which is why we call it the ``stochastic stepwise ensemble'' (or ST2E).
Explanations for why a more structured stochastic optimizer is desirable 
are given in Section~\ref{sec:tradeoff}. 

\subsection{Random {L}asso and stability selection}
\label{sec:otherVSEs}

The idea of using the ensemble approach for variable selection has started 
to catch up in recent years, for example, the ``random Lasso'' method 
\citep{rLasso} and the ``stability selection'' method \citep{stab-sel}. 
The latter consists of a general class of methods for structural 
estimation, including graphical modeling, but we shall 
limit our discussions here to the regression problem. 

For regression problems, these algorithms essentially give rise to 
different VSEs as we have defined them in Section~\ref{sec:VSE}, although 
they are {\em not} explicitly presented or labeled as ensemble algorithms 
by the authors. The ``random Lasso'' method, for 
instance, was presented to ``alleviate [various] limitations of [the] 
Lasso'' when dealing with highly correlated variables and ``large $p$, 
small $n$'' problems. 
The ``stability selection'' method, on the other hand, was presented to 
reduce the influence of regularization parameters and to provide ``finite 
sample familywise error control'' for the expected number of false 
discoveries. We shall see later that, while this method does provide very 
good error control for false discoveries, it appears to do so at the 
expense of missing true signals. 

Even though one can apply stability selection to different regression 
procedures, we will consider only its application to the Lasso 
\citep{lasso}.
At a very high level, both the random Lasso and the stability selection 
algorithms consist of running a randomized Lasso repeatedly 
on many bootstrap samples and aggregating the results. That is, they 
can both be regarded as VSEs. They differ on how the Lasso is 
randomized. For 
\citet{rLasso}, each Lasso is run with a randomly selected subset of 
variables; both the size of this subset and the $l_1$-regularization 
parameter $\lambda$ are selected by cross validation (CV) and fixed at the 
CV choices. For \citet{stab-sel}, all regularization parameters $\lambda 
\geq \lambda_{min}$ are considered; each Lasso is run by randomly scaling 
the regularization parameter for every regression coefficient; and the 
parameter $\lambda_{min}$ is chosen to control the expected number of 
false discoveries.

\subsection{Outline}

We proceed as follows. In Section~\ref{sec:main}, we describe a more 
structured approach to generate VSEs, the ST2 algorithm. We explain why 
the ST2 algorithm produces better ensembles than the genetic algorithm. We 
also describe how to set the tuning parameter in ST2. In 
Section~\ref{sec:ex}, we present a number of simulated and real-data 
examples, and compare ST2E with a number of other variable-selection 
algorithms, such as Lasso \citep{lasso} and its variations 
\citep[e.g.,][]{aLasso, reLasso}, LARS \citep{lars}, SCAD \citep{scad}, 
elastic net \citep{enet}, VISA \citep{visa}, and the two ensemble 
approaches mentioned above --- random Lasso \citep{rLasso} and stability 
selection \citep{stab-sel}. In Section~\ref{sec:discuss}, we make a 
few more general comments about the ensemble approach for variable 
selection, and present a simple extension of ST2E to tackle ``large $p$, 
small $n$'' problems.

\section{Stochastic stepwise selection}
\label{sec:main}

In this section, we describe a more structured stochastic search algorithm 
suitable for variable selection. In section~\ref{sec:tradeoff}, some 
explanations will be given as to why a more structured stochastic search 
is desirable. 

\subsection{The ST2 algorithm}
\label{sec:ST2}

Traditional stepwise regression combines forward and backward selection, 
alternating between forward and backward steps. In the forward step, each 
variable other than those already included is added to the current model, 
one at a time, and the one that can best improve the objective function, 
e.g., the AIC, is 
retained. In the backward step, each variable already included is deleted 
from the current model, one at a time, and the one that can best improve 
the objective function is discarded. The algorithm continues until no 
improvement can be made by either the forward or the backward step.

Instead of adding or deleting variables one at a time, ST2 adds or deletes 
a {\em group} of variables at a time, where the group size is {\em 
randomly} decided. In traditional stepwise, the group size is one and each 
candidate variable is assessed. When the group size is larger than one, as 
is often the case for ST2, the total number of variable groups can be 
quite large. Instead of evaluating all possible groups, only a {\em 
randomly} selected few are assessed and the best one chosen. 
Table~\ref{tab:stst} contains a detailed description of the ST2 algorithm.

\renewcommand{\baselinestretch}{1}
\begin{table}[hptb]
\centering
\caption{\label{tab:stst}%
The stochastic stepwise (STST or simply ST2) algorithm for variable selection.
}
\fbox{%
\begin{tabular}{p{0.9\textwidth}}

{\bf Repeat}

\benum

\item (Forward Step) Suppose $d$ variables, 
$\{x_{l_1}, x_{l_2}, ..., x_{l_d}\}$ 
are not in the current model. Initially, $d=p$.

\benum	

\item Determine the number of variables we want to add into the model, 
or the group size, say $g_f \leq d$. 
$^{\dagger}$

\item Determine the number of candidate groups that will be assessed, 
$k_f$. 
$^{\dagger}$

\item Generate $k_f$ candidate groups of size $g_f$, each by randomly 
choosing $g_f$ variables without replacement from the set,
$\{x_{l_1}, x_{l_2}, ..., x_{l_d}\}$. 

\item Assess each candidate group by adding it into the current model, one 
group at a time. The one that can best improve the objective function 
(e.g., the AIC) is added into the model.

\eenum

\item (Backward Step) Suppose $h$ variables, 
$\{x_{l_1}, x_{l_2}, ..., x_{l_h}\}$ 
are in the current model. 

\benum	

\item Determine the number of variables we want to delete from the 
model, or the group size, say $g_b \leq h$. 
$^{\dagger}$

\item Determine the number of candidate groups that will be assessed, 
$k_b$. 
$^{\dagger}$

\item Generate $k_b$ candidate groups of size $g_b$, each by randomly 
choosing $g_b$ variables without replacement from the set,
$\{x_{l_1}, x_{l_2}, ..., x_{l_h}\}$. 

\item Assess each candidate group by deleting it from the current model, 
one group at a time. The one that can best improve the objective function 
(e.g., the AIC) is deleted from the model.

\eenum

\eenum
{\bf Until} 

\bitem
\item[] no improvement can be made by either the forward or
the backward step.  
\eitem
\end{tabular}}
\newline
{\small $^{\dagger}$ Details for how the numbers $g_f, k_f, g_b, k_b$ are 
determined are given in Sections~\ref{sec:tunefunc} and 
\ref{sec:tuneparam}.}
\end{table}
\renewcommand{\baselinestretch}{1.25}

\subsection{Tuning functions}
\label{sec:tunefunc}

We now explain how the numbers $g_f, g_b, k_f, k_b$ (see 
Table~\ref{tab:stst}) are determined. Suppose we are doing a forward 
(backward) step and the potential predictors to be added (deleted) are 
$\{x_1, x_2, ..., x_m\}$. First, we need to determine $g$, the number of 
variables to add (delete). Intuitively, it seems reasonable that $g$ 
should depend on $m$, say
\[
 g=\phi_g(m).
\]
Second, given $g$, we have a total of $m \choose g$ possible groups of 
variables and need to determine $k$, the number of groups to assess. 
Intuitively, it also seems reasonable that $k$ should depend on $m \choose 
g$, say
\[ 
 k=\phi_k(m,g).
\]
We define the function $\phi_g$ as
\[
 \phi_g(m) \sim \mbox{Unif}(\Psi_m),
\]
where
\[
\Psi_{m} = \left\{ 
  1, 2, 3, ..., 
  \lfloor \lambda m + 0.5 \rfloor
  \right\},
\]
for some $0<\lambda<1$,
and $\lfloor x \rfloor$ is the largest integer not greater than $x$.
The function $\phi_k$ is defined as:  
\beqn
\label{eq:kappa}
 \phi_k(m,g) = 
 \left\lfloor 
 {m \choose g}^{1/\kappa} + 0.5
 \right\rfloor
\eeqn
for some $\kappa > 1$. 
We fix $\lambda=1/2$ and discuss how to choose the parameter
$\kappa$ later
(Section~\ref{sec:tuneparam}). Here, we first explain why the functions 
$\phi_g$ and $\phi_k$ are chosen to have these particular forms, and why 
we fix $\lambda=1/2$.

First, it is important that $g=\phi_g(m)$ is a stochastic and not a 
deterministic function. Consider the first forward step and the first 
backward step. Suppose there are $m=p=20$ potential predictors. If 
$\phi_g(m)$ is a deterministic function, say $\phi_g(20)=7$ and $\phi_g(7)=3$, 
then, for all individual paths, only models with $7$ predictors are 
assessed by the forward step and those with $7-3=4$ predictors are 
assessed by the backward step. Many models, such as those with 2 or 3 
predictors, are never assessed. Clearly, more flexibility is needed. 
According to our definition and with $\lambda=1/2$,  
$\phi_g(20) \sim 
\mbox{Unif}\{1,2,...,10\}$. In the first forward step, some paths 
may add 3 
variables, while others may add 1, 2, 4, 5, ..., or 10 variables.

Next, it makes sense that we should not add or delete too many variables 
in a {\em single} step. This is why we fix $\lambda=1/2$, so that at the 
most half of the available variables can be added or deleted. Notice that 
this restriction only applies to single search steps; it does not preclude 
the algorithm (which consists of many such steps) from selecting all 
the variables if necessary. 

Finally, we want the function $k = \phi_k(m, g)$ to be monotonically 
increasing 
in $m \choose g$ --- as more subsets become available, more candidate 
groups should be assessed in order to have a reasonable chance of finding 
an improvement.  However, we cannot afford to let this grow linearly in $m 
\choose g$ since it can be a very large number and that's why we use ${m 
\choose g}^{1/\kappa}$ for some $\kappa > 1$. 

\subsection{Tuning parameters}
\label{sec:tuneparam}

We now explain how to choose the tuning parameter, $\kappa$. For 
prediction ensembles, cross validation can be used to adjust various 
tuning parameters in order to maximize {\em prediction accuracy}, but this 
requires our validation data to contain the right answer. For 
variable-selection ensembles, however, this is not viable. We cannot 
empirically adjust its tuning parameter(s) to maximize {\em selection 
accuracy} because, no matter how we set data aside for validation, we 
won't know which variables are in the ``true model'' and which ones are 
not. For this reason, many researchers still use cross-validated 
prediction error to guide the choice of tuning parameters for their 
variable-selection procedure, 
but \citet{yuhong-ABIC} has clearly established that prediction 
accuracy and selection accuracy are fundamentally at odds with each other. 
In a more recent article,
\citet{statsci-explain-vs-predict} discusses various related issues from a 
less technical and more philosophical point of view. 
It must be stated without ambiguity that we are aiming for selection 
accuracy, not prediction accuracy. As such, cross validation is out of the 
question. Instead, we use ideas from \citet{randomForest} to help us 
specify our tuning parameter.

\subsubsection{Strength-diversity tradeoff}
\label{sec:tradeoff}

\def\cP{\mathcal{P}_{\theta}}

\citet{randomForest} studied prediction ensembles that he 
called random forests (RF):
\[
 \mbox{RF} = \{f(\myvec{x}; \theta_b): 
 \theta_b \overset{iid}{\sim} \cP, b = 1, 2, ..., B \}.
\]
For classification, $f(\myvec{x}; \theta_b)$ is a classifier completely 
parameterized by $\theta_b$, and the statement ``$\theta_b 
\overset{iid}{\sim} \cP$'' means that each 
$f(\cdot;\theta_b)$ is generated using an iid 
stochastic mechanism, $\cP$, e.g.,
bootstrap sampling \citep[e.g.,][]{bagging} and/or random subspaces 
\citep[e.g.,][]{tkho}. 

\citet{randomForest} proved that, for a random forest to be effective, 
individual members of the ensemble must be as good classifiers as 
possible, but they must also be as uncorrelated to each other as possible. 
In other words, a good classification ensemble is a {\em diverse} 
collection of {\em strong} classifiers.

Typically, the diversity of an ensemble can be increased by increasing 
$\var(\cP)$. But, unfortunately, this almost always reduces its strength. 
Therefore, it is important to use a stochastic mechanism $\cP$ with a 
``reasonable'' $\var(\cP)$. This basic principle has been noted elsewhere, 
too. For example, \citet{isle} described ensembles from an 
importance-sampling point of view. There, the corresponding notion of 
$\var(\cP)$ is simply the variance of the importance-sampling 
distribution, also referred to as the trial distribution. It is well-known 
in the importance-sampling literature that the variance of the trial 
distribution must be specified carefully \citep[e.g.,][Section 2.5]{liu}.

Although Breiman's theory of strength-diversity tradeoff is developed 
for prediction ensembles, some of these ideas can be borrowed for VSEs. In 
fact, this tradeoff explains why ST2E produces better variable-selection 
ensembles than PGA. Recall that PGA uses the genetic algorithm as the main 
stochastic mechanism to produce the ensemble, whereas ST2E uses a more 
structured ST2 algorithm (Section~\ref{sec:ST2}). Using Breiman's 
language, we can say that the more structured ST2 algorithm has a ``more 
reasonable'' variance than the genetic algorithm. It also allows us to 
exercise more control over the algorithm via the tuning parameter 
$\kappa$ in 
order to better balance the intricate tradeoff between strength and 
diversity.

\subsubsection{Computable measures of strength and diversity}
\label{sec:measures}

\def\se{\mathcal{S}(\myvec{E})}
\def\de{\mathcal{D}(\myvec{E})}

We now describe how to use the diversity-strength tradeoff to specify 
the tuning parameter, $\kappa$. 
Given $p$ potential covariates, let $\myvec{E}$ be a VSE of size $B$. The 
idea is to define {\em computable} measures of its diversity and strength, 
say $\de$ and $\se$, and choose $\kappa$ to balance the tradeoff 
between them.

Given a VSE, $\myvec{E}$, its diversity $\de$ can be measured by the 
average within-ensemble variation. For every variable $j$, there are $B$ 
independent measures of how important the variable is. The quantity 
\beqn
 v(j) = 
 \frac{1}{B-1} 
 \sum_{b=1}^B 
 \left[ \myvec{E}(b,j) - \frac{1}{B} \sum_{b=1}^B \myvec{E}(b,j)
 \right]^2
\eeqn
is the within-ensemble variance of these measures, and the quantity
\beqn
\label{eq:VSEdiv}
 \de = \frac{1}{p} \sum_{j=1}^p v(j),
\eeqn
is a measure of the average within-ensemble variation.

Let $F(\cdot)$ denote the objective function that each path of the VSE 
aims to optimize. We measure the mean strength of $\myvec{E}$ by the average 
percent improvement of the objective function over the null model, i.e.,
\beqn
\label{eq:VSEstrength}
 \se = \frac{1}{B} \sum_{b=1}^B 
 \frac{|F\left( \myvec{E}(b,\cdot) \right) - F_0|}{F_0}
\eeqn
where $F_0$ is the objective function evaluated at the null model, i.e., a 
model that does not contain any predictors. 

\subsubsection{Example}
\label{sec:ex-tune}

\begin{figure}[hptb]
\centering
\includegraphics[width=0.325\textwidth, angle=270]{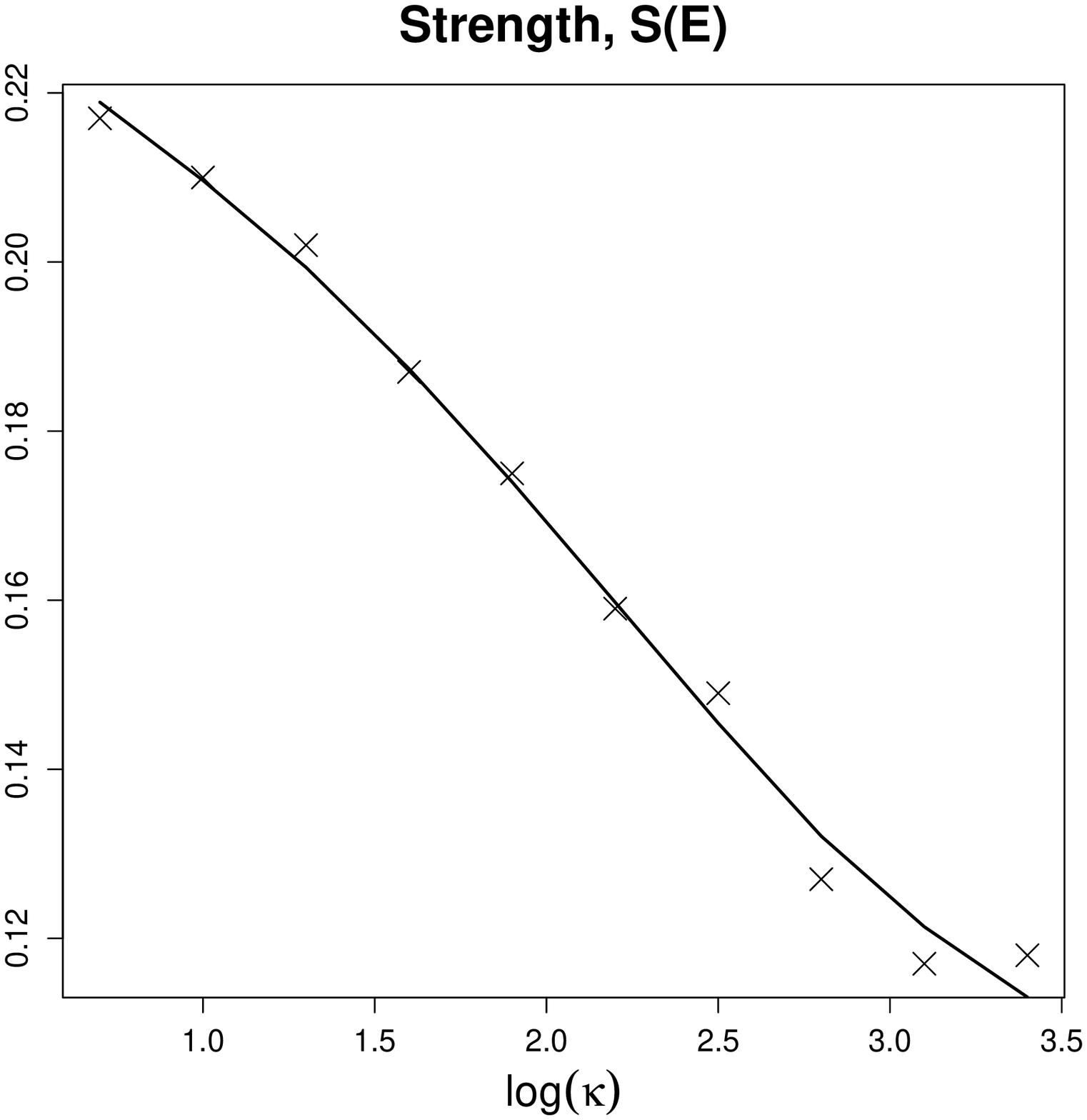}
\includegraphics[width=0.325\textwidth, angle=270]{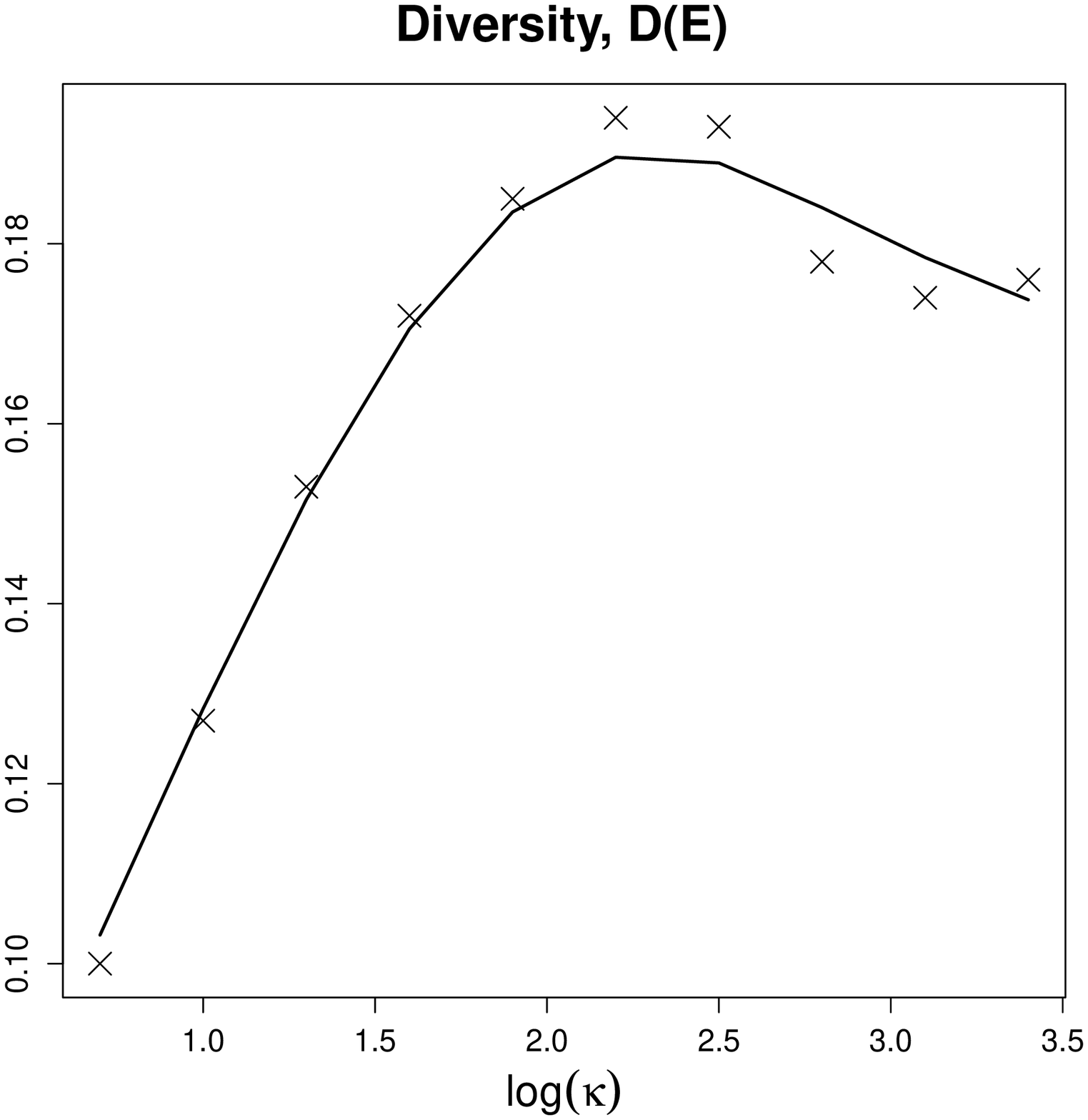}
\includegraphics[width=0.325\textwidth, angle=270]{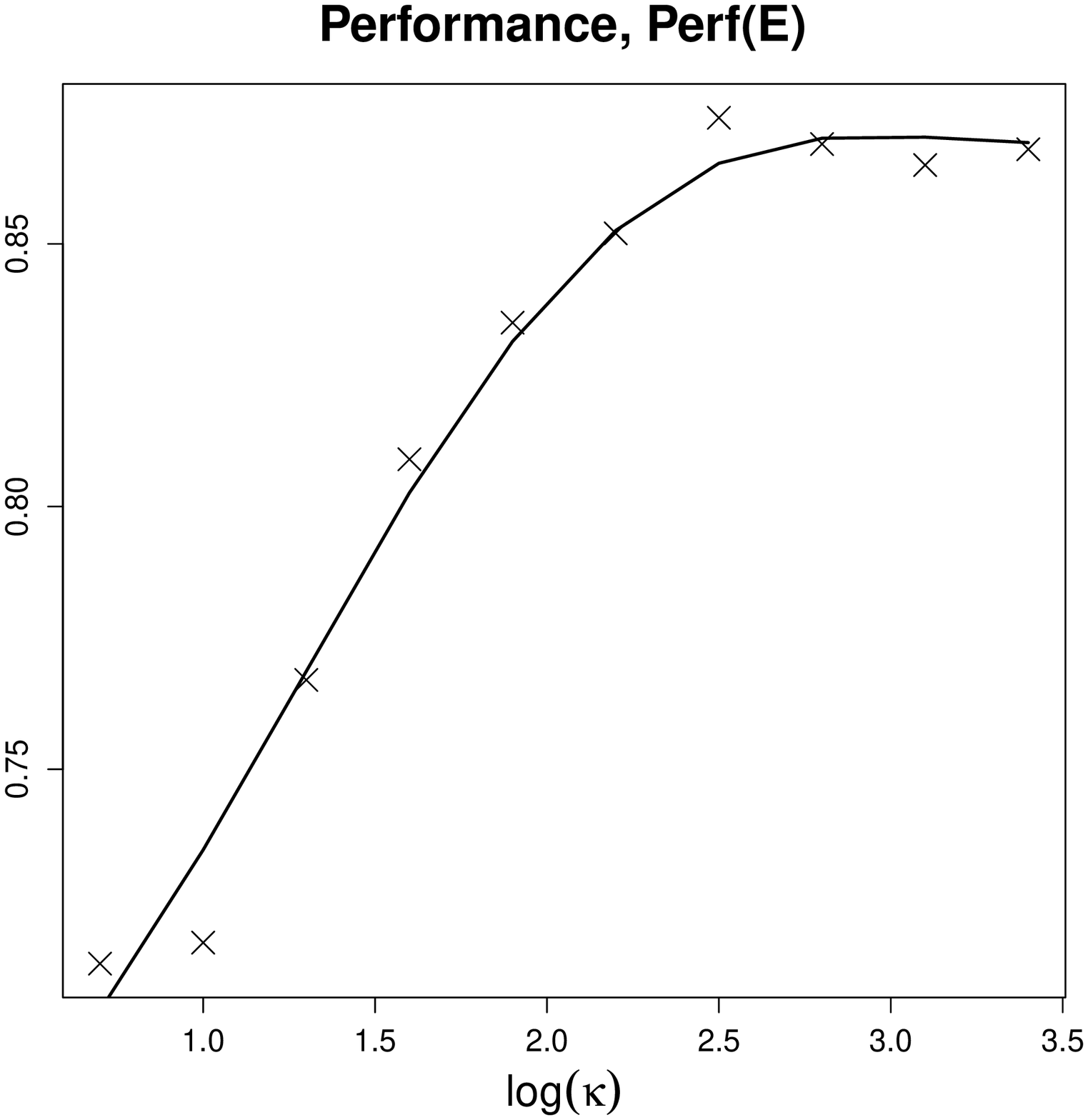}
\caption{\label{fig:ex1-tune}%
Illustration of how to select the tuning 
parameter, $\kappa$, in ST2E. Based on 
50 simulations of model (\ref{eq:ex0}) with $\alpha=1$.}
\end{figure}

Figure~\ref{fig:ex1-tune} shows an example of how our tuning strategy 
works. This is based on 50 simulations using equation~(\ref{eq:ex0}), 
while fixing $\alpha=1$. The quantities $\se$ and $\de$ are plotted 
against $\kappa$ (left and middle). For $\kappa$, the logarithmic scale is 
used. Variable-selection performance, measured here by
\beqnn
 \mbox{Perf}(\myvec{E}) = \mbox{ASF(Signal) - ASF(Noise)},
\eeqnn
where ``ASF'' stands for the average selection frequency, is 
also plotted (right). 

The behavior depicted here is fairly typical. We see from the left 
panel that $\se$ tends to 
decrease as we increase $\kappa$. This is because, when $\kappa$ is 
relatively small, the search conducted by steps 1(d) and 2(d) in the ST2 
algorithm (Table~\ref{tab:stst}) is relatively greedy; many candidate 
subsets (of a certain given size) are evaluated and the best one, chosen. 
This results in high strength. As $\kappa$ increases, the search becomes 
less greedy, which reduces strength.

On the other hand, the greedier the search, the higher the chance of 
finding the same subset. This explains why a small $\kappa$ tends to 
produce low diversity. As $\kappa$ increases and the search becomes less 
greedy, diversity starts to increase. But the parameter $\kappa$ controls 
not only the greediness but also the scope of the search. 
For example, it is easy to see from (\ref{eq:kappa}) that $\phi_k(m,g) 
\rightarrow 1$ as $\kappa \rightarrow \infty$. This means that, when 
$\kappa$ is very large, the scope of the search performed by steps 1(d) 
and 2(d) becomes quite limited, which also reduces diversity. This is why 
we see that, in the middle panel, the diversity measure $\de$ first 
increases but eventually decreases with $\kappa$.

Finally, we see from the right panel that, as $\de$ reaches its peak 
level, the variable-selection performance also starts to level off or 
drop. This means choosing $\kappa$ by looking for the peak in the $\de$ 
plot can be an effective strategy. This is what we use in all the 
experiments we report below.

Of course, we must emphasize that the measure $\mbox{Perf}(\myvec{E})$ and 
hence the right panel of Figure~\ref{fig:ex1-tune} are typically not 
available; 
they are only available for simulated examples where the true model is 
known. We include them here merely as a validation that it makes sense to 
choose $\kappa$ in such a way. In reality, one must rely on the plot of 
$\de$ alone to make the choice.

\subsection{Effect of sample size}
\label{sec:consistency}

Before we move on to examples and performance comparisons, we conduct a 
simple experiment to examine the performance of ST2E as the sample size 
$n$ increases. As in Section~\ref{sec:ex-tune}, we simulate from our 
motivating example, equation~(\ref{eq:ex0}), except we fix $\alpha=0.15$ 
this time. Using a small $\alpha$ creates a more difficult problem, which 
will allow us to see the effect of $n$ more clearly. For each $n = 50, 
100, 150, 250$ and $500$, we perform 100 simulations and record the 
average number of times the three types of variables (strong, weak, noise) 
are selected.  Table~\ref{tab:consistency} shows that ST2E behaves 
``reasonably'' in this respect. As $n$ increases, signal variables (both 
strong and weak) are selected with increasing probability, and the 
opposite is true for noise variables.

\renewcommand{\baselinestretch}{1}
\begin{table}[hptb]
\centering
\caption{\label{tab:consistency}%
Average number of 
times (out of 100) the three 
types of variables --- weak signal ($j=1$), strong signal ($j=2,3$), and 
noise ($j>3$) --- are selected by ST2E as the sample size $n$ varies. 
Same setting as the motivating example 
in Section~\ref{sec:ex0}, with $\alpha=0.15$.}
\fbox{%
\begin{tabular}{c|c|c|c} 
Sample Size 
& \multicolumn{1}{c|}{$\myvec{x}_j \in$ Weak Signal}
& \multicolumn{1}{c|}{$\myvec{x}_j \in$ Strong Signal}
& \multicolumn{1}{c}{$\myvec{x}_j \in$ Noise} \\
$(n)$ & \multicolumn{1}{c|}{$(j=1)$}
& \multicolumn{1}{c|}{$(j=2, 3)$}
& \multicolumn{1}{c}{$(j>3)$} \\ 
\hline
\phantom{1}50	&62	&\phantom{1}99	&17.12\\
100	&86	&100	&15.41\\
150	&89	&100	&13.18\\
250	&97	&100	&11.82\\
500	&99	&100	&11.53\\
\end{tabular}}
\end{table}
\renewcommand{\baselinestretch}{1.25}

\section{Examples}
\label{sec:ex}

We now present a few examples and compare the performance 
of ST2E with a number of other methods. For VSEs (ST2E, PGA, and stability 
selection), we use a size of $B=300$. To run stability selection, two 
tuning parameters are required, 
$\pi_{thr}$ and $\lambda_{min}$ --- see \citet{stab-sel} for details. The 
authors suggested fixing either one at a ``sensible'' value and choosing 
the other to control the expected number of false discoveries. We fix 
$\pi_{thr} = 0.6$, and report results for a range of $\lambda_{min}$ 
values.  

\subsection{A widely used benchmark} 
\label{sec:ex1}

First, we look at a widely used benchmark simulation. 
There are $p=8$ variables, $\myvec{x}_1, ..., 
\myvec{x}_8$, each generated from the 
standard normal. 
Furthermore, the variables are generated to be correlated, 
with $\rho(\myvec{x}_i,\myvec{x}_j) = 0.5^{|i-j|}$ for all $i \neq j$.
The response $\myvec{y}$ is generated by
\beqn
\label{eq:ex1}
 \myvec{y} = 3 \myvec{x}_1 + 1.5 \myvec{x}_2 + 2 \myvec{x}_5 
 + \sigma \gvec{\epsilon}
\eeqn
where $\gvec{\epsilon} \sim \normal{\myvec{0}}{\myvec{I}}$. 
That is, only three variables are true signals; the remaining five are 
noise. 
This benchmark was first used by \citet{lasso}, but it has been used by 
{\em almost} every major variable-selection paper published ever since.

For example, \citet[][Example 4.1]{scad} used this simulation to compare 
Lasso, the nonnegative garrote \citep{garrote}, hard thresholding 
\citep[see, e.g.,][]{thresh}, best subset regression 
\citep[e.g.,][]{miller:select}, 
and SCAD \citep{scad}. \citet[][Example 1]{rLasso} used this simulation to 
compare Lasso, adaptive Lasso \citep{aLasso}, elastic net \citep{enet}, 
relaxed Lasso \citep{reLasso}, VISA \citep{visa}, and random Lasso 
\citep{rLasso}.

\renewcommand{\baselinestretch}{1}
\begin{table}[hptb]
\centering
\caption{\label{tab:ex-scad}%
Widely-used benchmark (Section~\ref{sec:ex1}). Average number of zero 
coefficients for the noise group (oracle result = 5) and the signal 
group (oracle result = 0), based on 100 simulations. Results other than 
those for ST2E, PGA and stability selection are taken 
from \citet[][Table 1]{scad}. SCAD1 uses cross-validation to select the 
tuning parameter, while SCAD2 uses a fixed tuning parameter --- details 
are in \citet{scad}.} \fbox{%
\begin{tabular}{ll|cc}
& & \multicolumn{2}{c}{Avg.~No.~of 0 Coef.} \\
\multicolumn{2}{l|}{Method}  
& \multicolumn{1}{c}{$\myvec{x}_j \in $ Noise} 
& \multicolumn{1}{c}{$\myvec{x}_j \in $ Signal} \\
&  
& \multicolumn{1}{c}{$(j=3,4,6,7,8)$} 
& \multicolumn{1}{c}{$(j=1,2,5)$} \\
\hline
\multicolumn{2}{l|}{$n = 40, \sigma = 3$} \\
& SCAD1 & 4.20 & 0.21 \\
& SCAD2 & 4.31 & 0.27 \\
& LASSO & 3.53 & 0.07 \\
& Hard  & 4.09 & 0.19 \\
& Best subset & 4.50 & 0.35 \\
& Garrote     & 2.80 & 0.09 \\
& Oracle      & 5.00 & 0.00 \\
& ST2E  & 4.56 & 0.18 \\
& PGA   & 4.75 & 0.16 \\
& Stability selection \\
& ~~$\lambda_{min}=1.5$ & 4.96 & 1.03 \\
& ~~$\lambda_{min}=1.0$ & 4.86 & 0.58 \\
& ~~$\lambda_{min}=0.5$ & 4.54 & 0.18 \\
\multicolumn{2}{l|}{$n = 60, \sigma = 1$} \\
& SCAD1 & 4.37 & 0.00 \\
& SCAD2 & 4.42 & 0.00 \\
& LASSO & 3.56 & 0.00 \\
& Hard  & 4.02 & 0.00 \\
& Best subset & 4.73 & 0.00 \\
& Garrote     & 3.38 & 0.00 \\
& Oracle      & 5.00 & 0.00 \\
& ST2E  & 4.81 & 0.00 \\
& PGA   & 4.96 & 0.00 \\
& Stability selection \\
& ~~$\lambda_{min}=1.5$ & 5.00 & 0.21 \\
& ~~$\lambda_{min}=1.0$ & 5.00 & 0.01 \\
& ~~$\lambda_{min}=0.5$ & 4.95 & 0.00 \\
\end{tabular}}
\end{table}
\renewcommand{\baselinestretch}{1.25}

Results from \citet{scad} are replicated in Table~\ref{tab:ex-scad}, 
together with results from three VSEs --- ST2E, PGA, and stability 
selection. The ensemble algorithms are clearly among the most competitive. 
PGA and stability selection using a relatively large $\lambda_{min}$ are 
slightly better than ST2E at excluding noise variables, but, as will 
become clearer in Table~\ref{tab:ex-rLasso1}, they are also more likely 
than ST2E to miss true signals.

Results from \citet{rLasso} are replicated in Table~\ref{tab:ex-rLasso1}, 
together with results from ST2E, PGA, and stability selection. Here, the 
difficulty of PGA becomes clearer. It is better than ST2E in terms of 
excluding noise variables, but it also misses true signals more often. The 
same can be said about stability selection using a relatively large 
$\lambda_{min}$ value. It controls the number of false discoveries quite 
effectively, but we see that this will cause the method to behave poorly 
in terms of catching true signals. In order to improve its false negative 
rates, we must reduce $\lambda_{min}$, but this necessarily allows for 
more false discoveries. Random Lasso has the best ability to catch true 
signals, but to do so, it makes a large number of false discoveries 
at the same time.

\renewcommand{\baselinestretch}{1}
\begin{table}[hptb]
\centering
\caption{\label{tab:ex-rLasso1}%
Widely-used benchmark (Section~\ref{sec:ex1}). 
Minimal, median, and maximal number of times (out of 100 simulations) 
different types of 
variables (signal versus noise) are selected. 
Results other than those for ST2E, PGA and stability selection are taken 
from \citet[][Table 
2]{rLasso}.}
\fbox{%
\begin{tabular}{ll|rrr|rrr}
& & 
\multicolumn{3}{c|}{$\myvec{x}_j \in$ Signal} & 
\multicolumn{3}{c}{$\myvec{x}_j \in$ Noise} \\
& & 
\multicolumn{3}{c|}{$(j=1, 2, 5)$} & 
\multicolumn{3}{c}{$(j = 3, 4, 6, 7, 8)$} \\
\multicolumn{2}{l|}{Method} & Min & Median & Max & Min & Median & Max \\
\hline
\multicolumn{2}{l|}{$n=50, \sigma=1$} & & & \\
& Lasso           & 100 & 100 & 100 & 46 & 58 & 64 \\
& Adaptive Lasso  & 100 & 100 & 100 & 23 & 27 & 38 \\ 
& Elastic Net     & 100 & 100 & 100 & 46 & 59 & 64 \\ 
& Relaxed Lasso   & 100 & 100 & 100 & 10 & 15 & 19 \\ 
& VISA            & 100 & 100 & 100 & 11 & 17 & 20 \\
& Random Lasso    & 100 & 100 & 100 & 28 & 33 & 44 \\
& ST2E 
                  & 100 & 100 & 100 & 1 & 1 & 8 \\
& PGA             & 100  & 100 & 100 & 0  & 2  & 6 \\
& Stability selection & & & \\
& ~~$\lambda_{min}=1.5$ & 75 & 86 & 100 & 0 & 0 & 2 \\
& ~~$\lambda_{min}=1.0$ & 100 & 100 & 100 & 0 & 0 & 2 \\
& ~~$\lambda_{min}=0.5$ & 100 & 100 & 100 & 0 & 0 & 7 \\
\multicolumn{2}{l|}{$n=50, \sigma=3$} & & & \\
& Lasso           & 99  & 100 & 100 & 48 & 55 & 61 \\
& Adaptive Lasso  & 95  & 99  & 100 & 33 & 40 & 48 \\
& Elastic Net     & 100 & 100 & 100 & 44 & 55 & 69 \\
& Relaxed Lasso   & 93  & 100 & 100 & 11 & 18 & 21 \\
& VISA            & 97  & 100 & 100 & 15 & 21 & 24 \\
& Random Lasso    & 99  & 100 & 100 & 45 & 57 & 68 \\ 
& ST2E  
                  & 89  & 96  & 100 & 4  & 12 & 20 \\
& PGA             & 82  & 98  & 100  & 4  & 7  & 11\\
& Stability selection & & & \\
& ~~$\lambda_{min}=1.5$ & 59 & 64 & 100 & 0 & 0 & 3 \\
& ~~$\lambda_{min}=1.0$ & 81 & 83 & 100 & 0 & 2 & 9 \\
& ~~$\lambda_{min}=0.5$ & 90 & 98 & 100 & 4 & 8 & 22 \\
\multicolumn{2}{l|}{$n=50, \sigma=6$} & & & \\
& Lasso           & 76 & 85 & 99  & 47 & 49 & 53 \\
& Adaptive Lasso  & 62 & 76 & 96  & 32 & 36 & 38 \\
& Elastic Net     & 85 & 92 & 100 & 43 & 51 & 70 \\
& Relaxed Lasso   & 60 & 70 & 98  & 15 & 19 & 21 \\
& VISA            & 61 & 72 & 98  & 15 & 19 & 24 \\
& Random Lasso    & 92 & 94 & 100 & 40 & 48 & 58 \\
& ST2E   
                  & 68 & 69 & 96  & 9 & 13 & 21 \\
& PGA             & 54 & 76 & 94  & 9  & 14  & 16 \\
& Stability selection & & & \\
& ~~$\lambda_{min}=1.5$ & 40 & 41 & 83 & 0 & 4 & 8 \\
& ~~$\lambda_{min}=1.0$ & 59 & 61 & 92 & 4 & 8 & 18 \\
& ~~$\lambda_{min}=0.5$ & 76 & 84 & 100 & 30 & 42 & 50 \\
\end{tabular}}
\end{table}
\renewcommand{\baselinestretch}{1.25}

\subsection{Highly-correlated predictors}
\label{sec:ex2}

Next, we look at a simulation \citep[][Example 4]{rLasso} that is 
specifically created to study variable selection algorithms when the 
predictors are highly correlated {\em and} their coefficients have 
opposite signs. There are $p=40$ variables, $\myvec{x}_1, ..., 
\myvec{x}_{40}$, each generated from the standard normal. 
The response $\myvec{y}$ is generated by
\beqn
 \myvec{y} = 
  3 \myvec{x}_1 + 3 \myvec{x}_2  - 2 \myvec{x}_3 
+ 3 \myvec{x}_4 + 3 \myvec{x}_5  - 2 \myvec{x}_6 
+ \sigma \gvec{\epsilon}
\eeqn
where $\gvec{\epsilon} \sim \normal{\myvec{0}}{\myvec{I}}$ and $\sigma=6$. 
In addition, the 40 variables are generated to have the following 
(block diagonal) correlation structure:
\[
\left[
\begin{array}{ccc}
\myvec{C}_{3 \times 3}  & - & - \\
-  & \myvec{C}_{3\times 3} & - \\
-  & - & - \\
\end{array}
\right],
\quad\mbox{where}\quad
\myvec{C} = 
\left[
\begin{array}{lll}
1 & 0.9 & 0.9 \\
0.9 & 1 & 0.9 \\
0.9 & 0.9 & 1 \\
\end{array}
\right].
\]
That is, there are three groups of variables: 
$V_1 = \{1,2,3\}$,
$V_2 = \{4,5,6\}$, and
$V_3 = \{7,8,...,40\}$.
The first two groups, $V_1$ and $V_2$, are true signals, but,
within each of $V_1$ and $V_2$, the variables are highly correlated. 
There is no between-group correlation. 

\renewcommand{\baselinestretch}{1}
\begin{table}[hptb]
\centering
\caption{\label{tab:ex-rLasso4}%
Highly-correlated predictors (Section~\ref{sec:ex2}). 
Minimal, median, and maximal number of times (out of 100 simulations) 
different types of 
variables (signal versus noise) are selected. 
Results other than those for ST2E, PGA, and stability selection are taken 
from \citet[][Table 
2]{rLasso}.}
\fbox{%
\begin{tabular}{ll|rrr|rrr}
& & 
\multicolumn{3}{c|}{$\myvec{x}_j \in$ Signal} & 
\multicolumn{3}{c}{$\myvec{x}_j \in$ Noise} \\
& & 
\multicolumn{3}{c|}{$(j \leq 6)$} & 
\multicolumn{3}{c}{$(j > 6)$} \\
\multicolumn{2}{l|}{Method} & Min & Median & Max & Min & Median & Max \\
\hline
\multicolumn{2}{l|}{$n=50$} & & & \\
& Lasso           & 11 & 70 & 77 & 12 & 17 & 25 \\
& Adaptive Lasso  & 16 & 49 & 59 & 4  & 8  & 14 \\
& Elastic Net     & 63 & 92 & 96 & 9  & 17 & 23 \\
& Relaxed Lasso   & 4  & 63 & 70 & 0  & 4  & 9  \\ 
& VISA            & 4  & 62 & 73 & 1  & 3  & 8  \\
& Random Lasso    & 84 & 96 & 97 & 11 & 21 & 30 \\
& ST2E 
                  & 85 & 96 & 100 & 18 & 25 & 34 \\
& PGA             & 55 & 87 & 90 & 14  & 23 & 32 \\
& Stability selection  & & &\\
& ~~$\lambda_{min}=0.1$ & 1 & 35 & 42 & 1 & 5 & 13 \\
& ~~$\lambda_{min}=0.01$ & 1 & 37 & 45 & 7 & 13 & 22 \\
& ~~$\lambda_{min}=0.002$ & 1 & 40 & 52 & 31 & 42 & 54 \\
\multicolumn{2}{l|}{$n=100$} & & & \\
& Lasso           & 8  & 84 & 88 & 12 & 22 & 31 \\
& Adaptive Lasso  & 17 & 62 & 72 & 4  & 10 & 14 \\
& Elastic Net     & 70 & 98 & 99 & 7  & 14 & 21 \\
& Relaxed Lasso   & 3  & 75 & 84 & 1  & 3  & 8  \\
& VISA            & 3  & 76 & 85 & 1  & 4  & 9  \\
& Random Lasso    & 89 & 99 & 99 & 8  & 14 & 21 \\
& ST2E
                  & 93 & 100 & 100 & 14 & 21 & 27 \\
& PGA             & 40  & 85 & 92 & 13  & 22  & 33 \\
& Stability selection  & & &\\
& ~~$\lambda_{min}=0.5$ & 1 & 67 & 73 & 3 & 8 & 13 \\
& ~~$\lambda_{min}=0.3$ & 2 & 69 & 75 & 13 & 26 & 32 \\
& ~~$\lambda_{min}=0.2$ & 3 & 71 & 78 & 60 & 72 & 78 \\
\end{tabular}}
\end{table}
\renewcommand{\baselinestretch}{1.25}

Results from \citet{rLasso} are replicated in Table~\ref{tab:ex-rLasso4}, 
together with results from the ST2E, PGA, and stability selection. 
Clearly, ST2E and random Lasso are the top two performers for this 
problem. PGA has a much higher tendency than ST2E to miss true signals. 
Stability selection appears to have some difficulties here as well. If we 
choose $\lambda_{min}$ to control the false discoveries too aggressively, 
we start to lose signals quite badly. When its false discovery 
rate is as low as that of relaxed Lasso and VISA, for example, stability 
selection misses true signals much more often. On the other hand, even if  
$\lambda_{min}$ is relaxed to allow for many false discoveries in this 
case, stability selection still 
cannot seem to catch the signals as well as ST2E or random Lasso. This 
example clearly demonstrates that, when the predictors are highly 
correlated, the performances of PGA and stability selection can start to 
deteriorate quite significantly, whereas ST2E produces a much more robust 
VSE.

\subsection{Diabetes data}
\label{sec:diabetes}

Finally, we analyze the ``diabetes'' data set, which was used as the main 
example in the ``least angle regression'' (LAR) paper \citep{lars}. This 
is a real data set; a standardized version of the data set is available as 
part of the \verb!R! package, \verb!lars!. There are $n=442$ diabetes 
patients and $p=10$ variables, such as age, sex, body mass index, and so 
on. The response is a measure of disease progression.

\begin{figure}[hptb]
\centering
\includegraphics[width=0.485\textwidth]{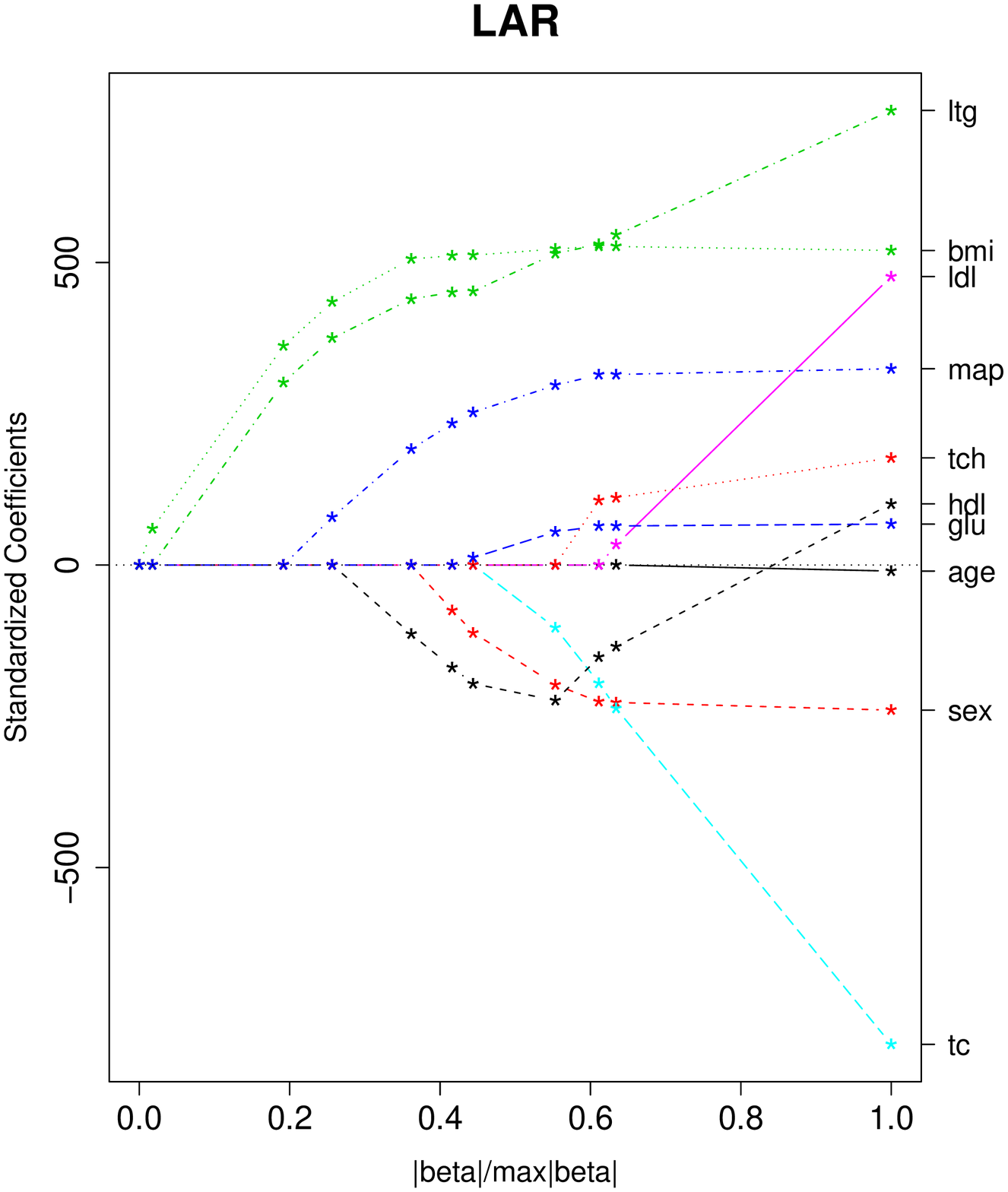}
\includegraphics[width=0.485\textwidth]{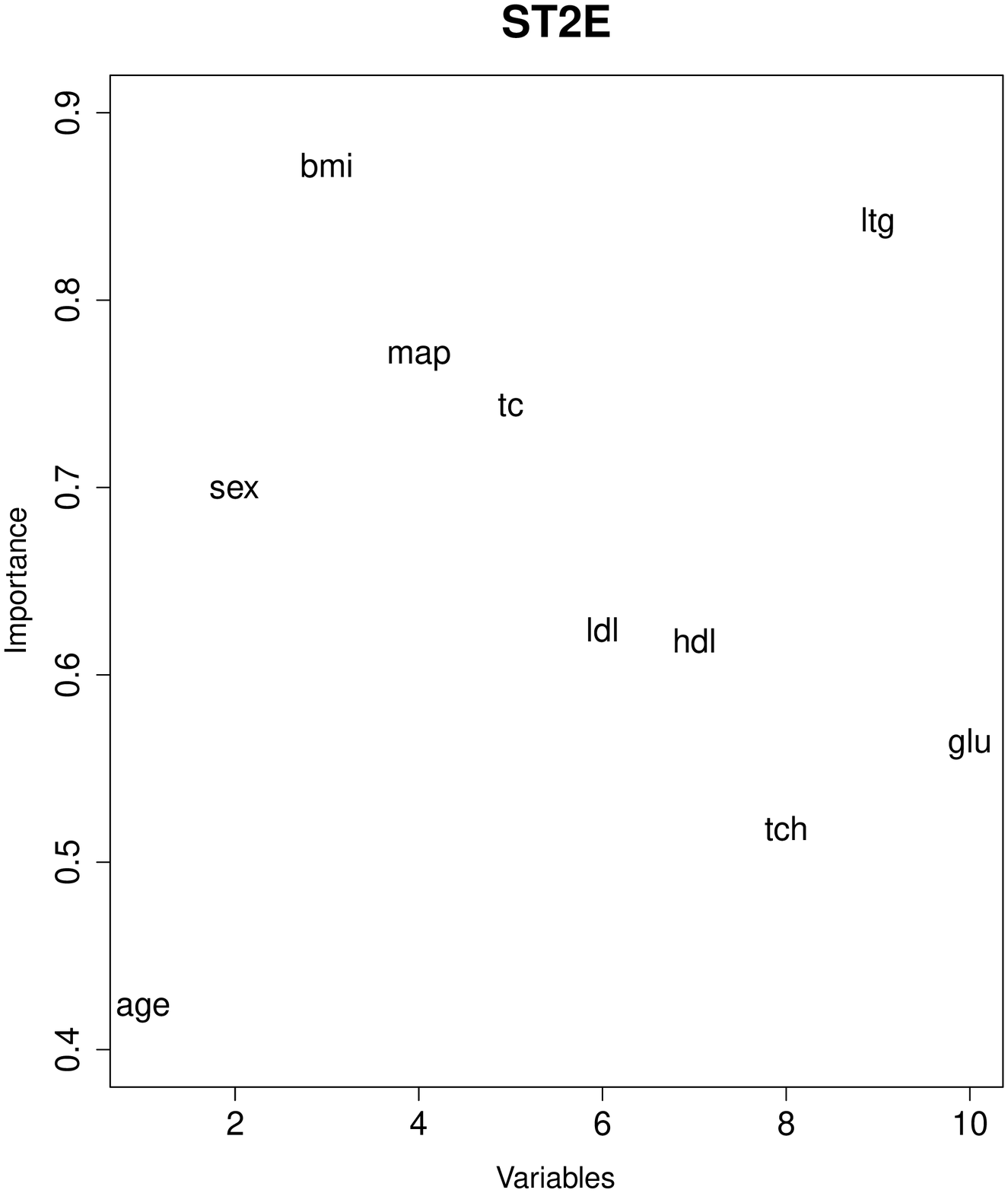}
\caption{\label{fig:diab}%
Diabetes data. Left: Results from least angle regression. Right: Results 
from ST2E.}
\end{figure}

Figure~\ref{fig:diab} shows results from both LAR and ST2E. For LAR, the 
entire solution paths are displayed for all the variables. As the penalty 
size decreases, the variables enter the model sequentially. The order in 
which they enter the model is listed in Table~\ref{tab:diab}. For ST2E, 
the variable importance measure (\ref{eq:rank}) are plotted for each 
variable. The order in which these variables are ranked by ST2E is also 
listed in Table~\ref{tab:diab}.

According to Table~\ref{tab:diab}, LAR and ST2E agree that ``bmi'', 
``ltg'', and ``map'' are the top three variables, whereas ``age'' is the 
least important variable. For the intermediate variables, LAR and ST2E 
seem to disagree on their relative importance. For example, the variable 
``ldl'' is the last one to be entered into the model by LAR before the 
variable ``age'', an indication that it is perhaps not an important 
variable, whereas ST2E ranks it in the middle.

Upon closer examination, however, we can see from Figure~\ref{fig:diab} 
that, once ``ldl'' is in the model, it actually gets a relatively large 
coefficient, larger than some of the other variables that were entered 
earlier --- this can almost certainly be attributed to ``ldl'' being 
highly correlated with these other variables. Therefore, there is good 
reason why ST2E does not rank ``ldl'' close to the bottom. Similar 
statements can be made for ``tc'', ``hdl'', and ``sex''.

\renewcommand{\baselinestretch}{1}
\begin{table}[hptb]
\centering
\caption{\label{tab:diab}%
Diabetes data. 
}
\fbox{%
\begin{tabular}{l|cccccccccc}
 & \multicolumn{10}{c}{Ordering and Ranking of Variables} \\
Method & \multicolumn{10}{c}{(top $\longrightarrow$ bottom)} \\
\hline
LAR  & bmi & ltg & map &hdl &sex &glu &tc &tch &ldl &age \\
ST2E & bmi & ltg & map &tc  &sex &ldl &hdl &glu &tch &age \\
\end{tabular}}
\end{table}
\renewcommand{\baselinestretch}{1.25}

\section{Discussions}
\label{sec:discuss}

Before we end, we would like discuss a few important issues. 

\subsection{Ranking versus thresholding}
\label{sec:rank-select}

The variable importance 
measure (\ref{eq:rank}) is a particularly nice feature for the ensemble 
approach. Using an ensemble approach, variable selection is performed in 
two steps: ranking and thresholding. We first rank the variables, e.g., by 
equation~(\ref{eq:rank}), and then use a thresholding rule to make the 
selection, e.g., equation~(\ref{eq:meanthresh}). 

As proponents of the ensemble approach, we are of the opinion that the 
task of ranking is the more fundamental of the two. From a 
decision-theoretic point of view, once the variables are ranked, the 
choice of the decision threshold has more to do with one's prior belief of 
how sparse the model is likely to be. 

We also think that variable {\em selection} per se is not quite the right 
objective, whereas variable {\em ranking} is. Imagine the problem of 
searching for a few biomarkers that are associated with a certain disease. 
What type of answer is more useful to the medical doctor? Telling her that 
you think it is biomarkers A, B, and C that are associated? Or giving her 
a ranked list of the biomarkers? Such a list is precisely what the 
ensemble approach aims to provide.

Therefore, we would have preferred not to introduce any thresholding rule 
at all. But, in order to compare with other methods in the literature, it 
is not enough to just rank the variables; we must make an active selection 
decision. In fact, because experiments are typically repeated multiple 
times, we must use a thresholding rule that is automatic, such as 
equation~(\ref{eq:meanthresh}). However, it is not our intention to take 
this thresholding rule too seriously; it is only introduced so that we 
could produce the same type of results as everybody else on various 
benchmark experiments. As far as we are concerned, the output of 
variable-selection ensembles is the variable importance measure 
(\ref{eq:rank}), which can be used to {\em rank} potential variables. 

\subsection{Stochastic generating mechanisms}
\label{sec:bigpic}

It is clear from our definition (Section~\ref{sec:VSE}) that there can be 
many ways to construct VSEs. To do so, we must employ a stochastic 
mechanism so that we can repeatedly perform traditional variable selection 
and obtain slightly different answers each time. One way to achieve this 
is to use a stochastic rather than a deterministic search algorithm to 
perform the selection, e.g., PGA \citep{pga} and ST2E (this paper). 
Another way is to perform the selection on bootstrap samples, e.g., random 
Lasso \citep{rLasso} and stability selection \citep{stab-sel}. According 
to our own empirical experiences (not reported in this paper), 
bootstrapping 
alone usually does not generate enough diversity within the ensemble to 
give satisfactory performance. This explains why extra randomization is 
employed by both the random Lasso and the stability selection methods; see 
Section~\ref{sec:otherVSEs}. Table~\ref{tab:compareVSEs} summarizes the 
stochastic mechanisms used to generate different VSEs. 

This leads us to a very natural question: what makes a good stochastic 
mechanism for generating VSEs? This question will likely take some time to 
answer; we certainly don't have an answer at the moment. What we have 
shown in this paper is the following: the ST2 algorithm appears to be 
quite an effective VSE-generating mechanism, and various ideas we have 
used to develop ST2E can help us think about this kind of questions in a 
more systematic manner.

\begin{table}
\centering
\caption{\label{tab:compareVSEs}%
Stochastic generating mechanisms for different VSEs.}
\fbox{%
\begin{tabular}{l|l}
VSE & \multicolumn{1}{c}{Generating Mechanism} \\
\hline
PGA & genetic algorithm \\
ST2E & stochastic stepwise search algorithm \\
Random Lasso & bootstrap + random subsets \\
Stability Selection & bootstrap + random scaling of regularization parameter 
\end{tabular}}
\end{table}

\subsection{``Large $p$, small $n$'' problems}
\label{sec:large-p-small-n}

We now describe a simple extension that allows ST2E to tackle ``large $p$, 
small $n$'' problems. To do so, we simply insert a pre-screening step 
before running each ST2 path by performing ``sure independence 
screening'' or SIS \citep{sis} on a bootstrap sample to pre-select $q < n$ 
variables, e.g., $q = n-1$. The ST2 algorithm is then applied to this 
subset of $q$ variables. Notice that, while this imposes an upper 
limit on how many variables {\em each} ST2 path can include, 
it by no means restricts the number of variables the ST2E ensemble can 
include as a whole. 

To test this strategy, we use another simulation from 
\citet[][Example 5]{rLasso} with $p=120$ and $n=50$, designed 
specifically as a test case for $p>n$ problems.  
The $120$ covariates are generated from a multivariate normal distribution 
with mean zero and covariance matrix,
\[
\left[
\begin{array}{cccc}
\myvec{\Sigma}_{30 \times 30} & - & - & - \\
- & \myvec{\Sigma}_{30 \times 30} & \myvec{J}_{30 \times 30} & - \\
- & \myvec{J}_{30 \times 30} & \myvec{\Sigma}_{30 \times 30} &- \\
- & - & - & \myvec{\Sigma}_{30 \times 30} \\
\end{array}
\right],
\]
where 
\[
\myvec{\Sigma}_{i,j} = 
\begin{cases}
1, & i=j \\
0.7, & i\neq j
\end{cases}
\quad\mbox{and}\quad
\myvec{J}_{i,j} = 0.2 \quad\forall\quad i, j.
\]
The first $60$ coefficients are generated from $\normal{3}{0.5}$ and then 
fixed for all simulations; the remaining $60$ coefficients are equal to 
zero. We are able to obtain and use exactly the same set of $60$ non-zero 
coefficients and the noise level for $\gvec{\epsilon}$ ($\sigma=50$) 
as those used by \citet{rLasso}.

\renewcommand{\baselinestretch}{1}
\begin{table}[hptb]
\centering
\caption{\label{tab:ex-rLasso5}%
A ``large $p$, small $n$'' problem (Section~\ref{sec:large-p-small-n}). 
Minimal, median, and maximal number of times (out of 100 simulations) 
different types of 
variables (signal versus noise) are selected. 
Results other than those for ST2E and stability selection are taken from 
\citet[][Table 
2]{rLasso}.}
\fbox{%
\begin{tabular}{l|rrr|rrr}
 & 
\multicolumn{3}{c|}{$\myvec{x}_j \in$ Signal} & 
\multicolumn{3}{c}{$\myvec{x}_j \in$ Noise} \\
 & 
\multicolumn{3}{c|}{$(j = 1, 2, ..., 60)$} & 
\multicolumn{3}{c}{$(j = 61, 62, ..., 120)$} \\
\multicolumn{1}{l|}{Method} & Min & Median & Max & Min & Median & Max \\
\hline
 Lasso           & 19 & 30 & 40 & 3  & 8  & 14 \\
 Adaptive Lasso  & 15 & 25 & 35 & 0  & 7  & 11 \\
 Elastic Net     & 40 & 50 & 61 & 1  & 5  & 8 \\
 Relaxed Lasso   & 14 & 23 & 34 & 0  & 3  & 8  \\ 
 VISA            & 16 & 27 & 35 & 0  & 2  & 8  \\
 Random Lasso    & 76 & 86 & 95 & 18 & 29 & 38 \\
 ST2E (with SIS)
                  & 81 & 88 & 95 & 0  & 1  & 5 \\
 Stability selection  & & &\\
 ~~$\lambda_{min}=10^{-2}$ & 4  & 22 & 52 & 0 & 3 & 10 \\
 ~~$\lambda_{min}=10^{-4}$ & 13 & 40 & 80 & 1 & 6 & 25 \\
 ~~$\lambda_{min}=10^{-5}$ & 73 & 95 & 100 & 16 & 38 & 80 \\
\end{tabular}}
\end{table}

Table~\ref{tab:ex-rLasso5} shows the results. Again, random Lasso is 
competitive in terms of catching the signals, but it makes a large number 
of false discoveries. Stability selection has the same difficulty as 
before --- it misses signals when large values of $\lambda_{min}$ are used 
to regulate false discoveries, and its ability to catch the signals can 
match that of ST2E and random Lasso only if $\lambda_{min}$ is set to be 
so low as to tolerate a very large number of false discoveries.
 
Readers will find that the 
performance of ST2E (with SIS) on this $p>n$ example is quite remarkable. 
This is because the extra pre-selection step really makes it an entirely 
different VSE altogether, since the stochastic generating mechanism (see 
Section~\ref{sec:bigpic} and Table~\ref{tab:compareVSEs}) has 
fundamentally changed. 
Intuitively, it is easy to see that SIS can improve ensemble strength by 
screening out noise variables, while doing so on bootstrap samples can 
further enhance ensemble diversity. This reinforces the points we have 
made earlier in Section~\ref{sec:bigpic} --- the question of how to design 
and/or characterize a good stochastic mechanism for generating VSEs is a 
very intriguing one indeed, and thinking about the strength and diversity 
of the ensemble often gives us some useful insights.

\subsection{False positives versus false negatives}

Another striking phenomenon that we can observe from all our experiments 
is the extremely delicate balance between false positive and false 
negative rates in variable selection problems. It is very hard to reduce 
one without significantly 
affecting the other. For VSEs, the underlying stochastic generating 
mechanism is, again, critical. Among the four VSEs listed in 
Table~\ref{tab:compareVSEs}, PGA seems to ``care'' more about false 
positives, whereas random Lasso appears to ``care'' more about false 
negatives. Stability selection allows users to control the number of false 
positives through a tuning parameter, but, as our experiments have shown, 
aggressive control of false positive rates necessarily leads to very poor 
false negative rates, and vice versa.

Overall, ST2E appears to balance the two objectives nicely, and this is 
precisely why we think it is a valuable practical algorithm. But, of 
course, there is no reason to believe why one cannot find another 
generating mechanism to produce a VSE that will balance the two objectives 
even better. However, as we've alluded to earlier, the more interesting 
question is not whether we can find another mechanism, but 
how we can know that we have found a good one. 
This we leave to future research.

\section{Summary}
\label{sec:summary}

We are now ready to summarize the main contributions of this paper. First, 
we gave a formal and general description of the ensemble approach for 
variable selection. Next, we pointed out that Breiman's theory for 
prediction ensembles --- in particular, the tradeoff between diversity and 
strength --- is useful in guiding the development of variable-selection 
ensembles as well. Finally, we used a more structured stochastic 
mechanism, the ST2 algorithm, to construct a better variable-selection 
ensemble, ST2E, which we demonstrated to be more robust than other VSEs, 
and competitive against many state-of-the-art algorithms.

\bibliography{/u/m3zhu/mzstat}
\end{document}